\documentclass[a4paper,UKenglish]{oasics}
 
\usepackage{microtype}
\usepackage{multirow}
\usepackage{subfig}
\usepackage{wrapfig}
\usepackage{pifont} 
\usepackage{color}
\usepackage{amsmath}
\usepackage{xspace} 
\usepackage[usenames,dvipsnames,svgnames,table]{xcolor}
\makeatletter
\newcommand{\thickhline}{
    \noalign {\ifnum 0=`}\fi \hrule height 1pt
    \futurelet \reserved@a \@xhline
}
\newcolumntype{"}{@{\hskip\tabcolsep\vrule width 1pt\hskip\tabcolsep}}
\makeatother 
\newcommand{\cmark}{{\color{DarkGreen}\ding{51}}}
\newcommand{\xmark}{{\color{red}\ding{55}}} 
\newcommand{\etal}{\mbox{\textit{et al.}}\xspace}
\newcommand{\kary}{\mbox{$k$-ary}\xspace}

\title{Faster Concurrent Range Queries with Contention Adapting Search Trees Using Immutable Data}
\titlerunning{Faster Concurrent Range Queries in CA Trees Using Immutable Data Structures} 

\author[1]{Kjell Winblad}
\affil[1]{Department of Information Technology, Uppsala University, Sweden\\
  \texttt{kjell.winblad@it.uu.se}}
\authorrunning{K. Winblad}

\Copyright{Kjell Winblad}

\subjclass{D.2.8 Performance measures, E.1 Trees, H.2.4 Concurrency}
\keywords{linearizability, concurrent data structures, treap}

\serieslogo{}
\volumeinfo
  {}
  {0}
  {To appear in 2017 Imperial College Computing Student Workshop (ICCSW 2017)}
  {3}
  {1}
  {1}
\EventShortName{}
\DOI{10.4230/OASIcs.xxx.yyy.p}

\begin{document}
 
\maketitle

\begin{abstract}
  The need for scalable concurrent ordered set data structures with linearizable range query support is increasing due to the rise of multicore computers, data processing platforms and in-memory databases.
  This paper presents a new concurrent ordered set with linearizable range query support.
  The new data structure is based on the contention adapting search tree and an immutable data structure.
  Experimental results show that the new data structure is as much as three times faster compared to related data structures.
  The data structure scales well due to its ability to adapt the sizes of its immutable parts to the contention level and the sizes of the range queries.
\end{abstract} 

\section{Introduction\label{sec:intro}}
The use of concurrent ordered set data structures\footnote{A concurrent ordered set data structure represents a set of items that can be manipulated concurrently by several threads and where an item consists of an ordered key and optionally some additional data.} with support for linearizable\footnote{A linearizable operation appears to happen instantly between the operation's invocation and return~\cite{HerlihyLinearizability}.} range queries\footnote{A range query operation returns all items with keys within the given range (specified by two keys).} is increasing as multicores are becoming more readily available and due to the rise of big scale data processing platforms and in-memory databases such as Google's F1~\cite{ShuteF1} and Yahoo's Flurry~\cite{Furry}.
Both of these require set data structures with fast updates\footnote{An update operation is an insert operation or a remove operation where the former inserts an item (replacing an existing item if one with an equal key already exists) and the latter removes an item with the given key if such an item exists.} to store incoming data while concurrently serving (typically large) linearizable range queries for analytics~\cite{BasinKiWi}.
Although there are many concurrent set data structures (e.g.~\cite{shalevExtendableLockFreeHash, michaelHashTable, ProkopecCtrie}) and ordered set data structures (e.g.~\cite{fraser2004practical,NatarajanFastLockFree,NBKASearchTree}), there are only a few concurrent data structures with efficient linearizable range queries~\cite{RangeQKArySeachTree,AvniLeaplist,CFRangeQueries,CATreeLCPC,ChatterjeeLFRangeQueries,BasinKiWi}.

This paper proposes a new concurrent ordered set data structure that internally makes use of an immutable data structure.
The difference between an immutable data structure and its mutable counterpart is that the immutable data structure's update operations do not modify the given data structure instance in-place but instead return a new version, leaving the input instance intact.
For many data structures, e.g. binary search trees, the operations of the immutable version are asymptotically as efficient as in its mutable counterpart~\cite{okasaki1999purely}.
As an example, the insert operation of an immutable balanced binary search tree only needs to make a copy of the nodes on the path to the inserted node, which only consists of $\mathcal{O}(\log{}n)$ nodes, where $n$ is the number of items that are stored in the search tree. 

One can derive a concurrent ordered set data structure with linearizable range query support from a single mutable reference to an immutable data structure~\cite{HerlihyUniMethod}.
A lookup or a range query simply performs the operation in the referenced immutable data structure.
An update operation repeatedly tries to update the reference using an atomic compare-and-swap operation~\cite{HerlihyUniMethod} until the update succeeds.
Unfortunately, this coarse-grained approach does not scale when concurrent updates are common due to the scalability bottleneck that exists in the updating of the shared reference.
Instead, several data structures~\cite{RangeQKArySeachTree,AvniLeaplist} use immutable parts that can store a fixed number of items to shorten the time range queries need to spend reading shared mutable data.
This fine-grained approach can be efficient when it is possible to fine-tune the size of the data structure's immutable parts to fit the sizes of the range queries (the number of items within the range) and the contention level.
However, the fine-grained approach does not work well when the access pattern is unknown or differs in different parts of the data structure.

The \emph{main contribution} of this paper is a new concurrent ordered set data structure with linearizable range query support that solves the problems with the coarse-grained and fine-grained approaches described above by dynamically changing the sizes of its immutable parts to fit the workload at hand.
The new data structure is based on the contention adapting search tree (CA tree)~\cite{CATreeISPDC, CATreeLCPC} and an immutable data structure.
Previous results~\cite{CATreeLCPC} show that CA trees using mutable data structures provide good scalability in scenarios with short range queries.
However, previous CA~tree variants' scalability for large range queries is limited as their range queries lock out other threads from large portions of the data structure for a time period whose length is proportional to the number of items with keys in the given range.
The new CA tree variant eliminates this problem by utilizing an immutable data structure.
As is shown in this work, the new CA tree variant's ability to reduce the lock holding times does not only make its scalability substantially better compared to the old CA tree variants but also much better than the other recently proposed data structures with linearizable range query support.

This paper starts with a high-level description of CA trees (Section~\ref{sec:catreedesc}).
The new CA tree variant and its implementation are described in Sections~\ref{immopt} and~\ref{immoptimpl}.
Analytical and experimental comparisons with related data structures are given in Sections~\ref{sec:related} and~\ref{sec:eval}.
The paper finishes with a conclusion (Section~\ref{sec:conc}).

\section{High-Level Description of Contention Adapting Search Trees\label{sec:catreedesc}}
\begin{wrapfigure}[13]{r}{.315\textwidth}
  \centering
  \includegraphics[width=.315\textwidth]{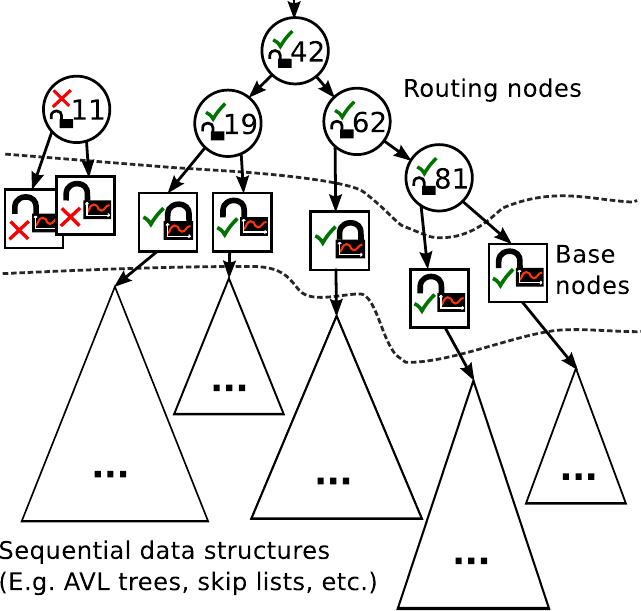}
  \caption{The structure of a CA tree. Numbers denote keys.}
  \label{fig:ca_structure}
\end{wrapfigure}
\begin{figure}[t]
  \newcommand{\graphheight}{.19\textheight}
  \newcommand{\bsp}{\hspace*{20pt}}
  \centering 
  \subfloat[Initial CA tree]{
    \includegraphics[height=\graphheight]{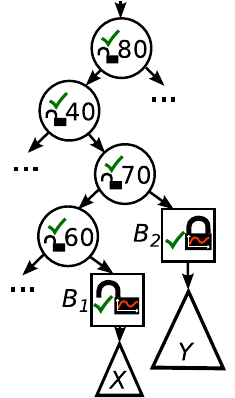}
    \label{fig:initial}
  } \bsp
  \subfloat[CA tree after a split]{
    \includegraphics[height=\graphheight]{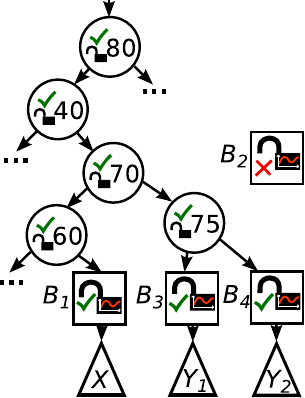}
    \label{fig:split}
  } \bsp
  \subfloat[CA tree after a join]{
    \includegraphics[height=\graphheight]{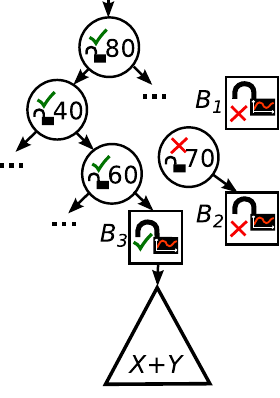}
    \label{fig:join}
  }
  \caption{Effect of the split and join operations on the CA tree of \figurename~\ref{fig:initial}.}
  \label{fig:split+join}
\end{figure}

CA trees are structured as  depicted in Figure~\ref{fig:ca_structure}.
The items that are stored in a CA tree are located in sequential data structures under the base nodes (see Figure~\ref{fig:ca_structure}).
To efficiently find a specific item in a CA tree, the search is directed by the keys in the routing nodes.
All items stored under the left branch of a routing node have keys that are less than the key of the routing node and all items stored under the right branch have keys that are greater than or equal to the key in the routing node.
The sequential data structures are protected from concurrent accesses by locks in the base nodes.
A base node lock has a statistics counter which is incremented when a thread needs to wait to acquire the lock and decremented when no waiting is required.
If the statistics counter in a base node reaches a certain threshold, the items stored in the base node are split between two new base nodes to reduce the contention; see~\figurename~\ref{fig:initial} and \figurename~\ref{fig:split}.
In the reverse direction, if the statistics counter in a base node reaches the threshold for low contention adaption, the items in the base node and a neighbor base node are joined into one new base node; see~\figurename~\ref{fig:initial} and \figurename~\ref{fig:join}.
Base nodes also have a valid flag (depicted by \cmark~and \xmark) which is used to indicate if a base node is in the CA tree or if it has been removed.
Operations that end up in an invalid (\xmark) base node need to retry the search until they end up in a valid (\cmark) base node.
Routing nodes also have a valid flag and a lock that are only used rarely during the low contention optimizing join.
Range queries are performed in CA trees by first finding and locking the base node containing the first key in the range and then traversing and locking subsequent base nodes until a base node containing a key which is equal to or greater than the largest key in the range is found.  

An optimization that has been shown to greatly enhance performance of read operations (lookup and range query) is to let read operations optimistically attempt to do their operation without writing to shared memory~\cite{CATreeISPDC, CATreeLCPC}.
This can be done by using a \emph{sequence lock}~\cite{lameter2005effective} as base node lock.
A sequence lock has an operation to read a sequence number from the lock.
If a thread gets the same even sequence number from two calls of this operation, then the sequence lock guarantees that the lock has not been acquired between the two calls.
An optimistic attempt of a read operation first scans the sequence numbers and checks the valid flags of the base nodes that the operation needs to read data from, and then performs the operation, after which the sequence numbers from the locks have to be checked again to make sure that the sequence numbers match the previously read sequence numbers.
If the optimistic attempt fails, the operation is done by acquiring the base node locks in read-mode (several read-mode lock holders can hold the lock at the same time).

The reader is referred to the earlier papers on CA trees~\cite{CATreeISPDC,CATreeLCPC} for a detailed description including pseudo-code and arguments that their operations provide linearizability, deadlock freedom, and livelock freedom.

\section{CA Tree Optimization Enabled by Immutable Data Structures\label{immopt}}
By using a mutable reference to an immutable data structure as a CA tree's sequential data structure,
it is straightforward to reduce the amount of time that read operations spend on reading shared mutable data.
Assuming that the CA tree's sequential data structure component is implemented with a mutable reference to an immutable data structure,
a lookup or range query operation only needs to copy the values of the references that are needed by the operation while traversing shared mutable data.
The immutable data structures referenced by the copied references are then traversed after the base node locks have been unlocked (or after the second sequence number scan).
Especially for range queries, this optimization can give a large reduction of the amount of time which is spent on reading shared mutable data as the time that range queries need to spend on traversing the sequential data structures is at least linear in the number of items in the range.
With the optimization, range queries may only need to traverse a few base nodes (one in the best case) while reading shared mutable data even when the number of items in the range is large. 

It is straightforward to see that this optimization does not jeopardize correctness as the result of a read operation would be the same if the traversal of the sequential data structures happened instantly at the linearization point (due to the immutability of the data structures referenced by the copied references).

\section{The Implementation of the Optimized CA Tree\label{immoptimpl}}
To experimentally evaluate the optimization described in the previous section, a CA tree using a mutable reference to an immutable treap~\cite{Seidel1996} as its sequential data structure has been implemented in Java.
A treap is a self-balancing binary search tree with expected time complexity of $\mathcal{O}(\log{}n)$ for insert, remove and lookup and an expected time complexity of $\mathcal{O}(\log{}n +r)$ for range queries, where $n$ is the number of items stored in the data structure and $r$ is the number of items in the range.
The treap also has efficient split and join operations~\cite{Seidel1996} which is important for the CA tree's low and high contention adaptions~\cite{CATreeISPDC}.
To facilitate cache friendly range queries, the treap implementation stores all items in fat leaf nodes containing arrays that can store up to 64 items.

One heuristic, that CA trees use to reduce the time that future similar range queries need to spend on traversing base nodes, is to decrement the contention statistics counters in the locks of base nodes needed by a range query, if more than one base node is needed; cf.~\cite{CATreeLCPC}.
With the optimization described in the previous section in place, the portion of a range query that is spent on traversing shared mutable data is even more affected by the number of base nodes that the range query needs to access than without the optimization.
The reason for this is that the optimization moves the traversal of the sequential data structures from within the period that is spent on reading shared mutable data to after this period. 
It therefore makes sense to decrement the contention statistics counters with a larger value, when the optimization is used, as the potential benefit for similar range queries is larger than without the optimization.
Indeed, experiments show that changing the heuristic to decrement by the value 100 instead of the value one (which is used by the old CA tree implementations) gives significantly better performance in scenarios with large range queries.
Except for the change of the value used to decrement the statistics counters in the described heuristic, the immutable treap version of the CA tree has the same constants and thresholds for low and high contention adaptions as described in the previous work~\cite{CATreeLCPC}.

\section{Related Work\label{sec:related}}
The CA tree is the only one of the previously proposed approaches for linearizable range queries~\cite{bronson2010practical,RangeQKArySeachTree,AvniLeaplist,CFRangeQueries,ChatterjeeLFRangeQueries,BasinKiWi} that dynamically changes the synchronization granularity to optimize for the conflicting requirements of range queries of different sizes and single-key operations~\cite{CATreeLCPC}.

The \emph{SnapTree} by Bronson~\etal~\cite{bronson2010practical} has an efficient linearizable clone operation that returns a copy of the data structure from which a range query operation can easily be derived.
\emph{SnapTree}'s clone operation waits for active update operations and forces subsequent update operations to copy nodes lazily before node modifications so that the copy is not modified.
The behavior of \emph{SnapTree} resembles the behavior of the data structure implemented from a single mutable reference to an immutable data structure (that is discussed in Section~\ref{sec:intro}) when range queries are common.
The \emph{SnapTree} can thus serve as an example of the coarse-grained approach for doing range queries.

The lock-free \kary{} search tree is an unbalanced external search tree with up to $k$ keys stored in every node~\cite{NBKASearchTree}.
Range queries in \kary{} search trees are performed by doing a read scan and a validation scan of the immutable leaf nodes containing items in the range~\cite{RangeQKArySeachTree}.
The range query operation needs to retry if the validation scan fails.
The \kary{} is an example of the fine-grained approach discussed in Section~\ref{sec:intro}.
Another example of this fine-grained approach based on software transactional memory is the Leaplist~\cite{AvniLeaplist}.
  
Chatterjee has proposed a general method for doing range queries in lock-free ordered set data structures~\cite{ChatterjeeLFRangeQueries} based on the work by Erez and Shahar~\cite{Petrank2013}.
Unfortunately, the scalability of Chatterjee's method suffers from the global sequential hot spot in the list of range-collector objects that all range queries have to write to in the worst case. 

The \emph{KiWi} data structure by Basin \etal~\cite{BasinKiWi} supports wait-free range queries and lookup operations as well as lock-free update operations.
Update operations help range queries by storing multiple versions of inserted items when it is needed for the range queries.
Similarly to Robertson's data structure~\cite{CFRangeQueries}, \emph{KiWi}'s range queries atomically increment a global version counter which is used by update operations to decide if storing an additional version is necessary.
\emph{KiWi}'s global version number counter is bound to become a scalability bottleneck with a high enough level of parallelism.
Similarly to the treap based CA tree, \emph{KiWi} tries to improve cache locality by storing items in arrays that can store up to $k$ items.

A fundamental difference between the other efficient methods for range queries in ordered set data structures and the optimized CA tree is the time spent by range queries reading shared mutable data and thus the time in which conflicts with update operations can happen.
In the other methods~\cite{RangeQKArySeachTree,AvniLeaplist,CFRangeQueries,CATreeLCPC,ChatterjeeLFRangeQueries,BasinKiWi}, this time is at least linear in the number of items inside the range while the optimized CA tree can do much better as is described in Section~\ref{immopt}.

\section{Evaluation\label{sec:eval}}
We now experimentally evaluate the optimized CA tree implementation using the immutable treap described in Section~\ref{immoptimpl} (\emph{Im-Tr-CA}).
\emph{Im-Tr-CA} will be compared to the recently proposed methods for doing linearizable range queries in ordered sets: \emph{SnapTree}~\cite{bronson2010practical}, \kary{}~\cite{RangeQKArySeachTree}, Chatterjee's method applied to a lock-free skiplist~\cite{ChatterjeeLFRangeQueries} (\emph{ChatterjeeSL}), \emph{KiWi}~\cite{BasinKiWi} and the two CA tree variants that use a mutable skiplist with fat nodes (\emph{SL-CA}) and a mutable AVL tree (\mbox{\emph{AVL-CA}}) as sequential data structures~\cite{CATreeLCPC}.
The lock-free skiplist, called ConcurrentSkipListMap, from the Java library that only supports range queries that are not linearizable (\emph{NonAtomicSL}) is also included in the comparison.
All data structures were provided by their respective authors and are implemented in Java.
The maximum number of items in the nodes ($k$) is set to 64 for \kary, \emph{Im-Tr-CA} and \emph{SL-CA} as this value has previously been shown to give good results~\cite{RangeQKArySeachTree}.
\emph{KiWi}'s constants are set as described in the \emph{KiWi} paper~\cite{BasinKiWi}.

The benchmarks were run on a machine with four Intel(R) Xeon(R) E5-4650 CPUs (2.70GHz each with eight cores and hyperthreading, giving a total of 32 actual and 64 logical cores), turbo boost turned off, 128GB of~RAM, running Linux 3.16.0-4-amd64 and Oracle JVM~1.8.0\_131 (with the JVM flags -Xmx8g -Xms8g -XX:+UseCondCardMark -server -d64).
Each data point comes from the average of three measurements runs of 10 seconds each that were preceded by 3 warm up runs, also of 10 seconds each.
The purpose of the warm up runs is to give the just-in-time compiler enough time to compile the code.
Error bars showing the minimum and maximum measurements are displayed when they are large enough to be seen.

The keys for the operations lookup, insert and remove as well as the starting keys for range queries are randomly generated from a range of size $S$.
The data structure is pre-filled before the start of each benchmark run by performing $S/2$ random insert operations.
In all experiments presented in the main part of this paper $S = 10^6$.
The interested reader can find results for $S = 10^5$ and $S = 10^7$ in the appendix of this paper.
Range queries calculate the sum of the items in the range and the number of items in the range.
As a sanity check, the average number of items that are traversed per range query is calculated and checked against the expected value.
\begin{figure}[b]
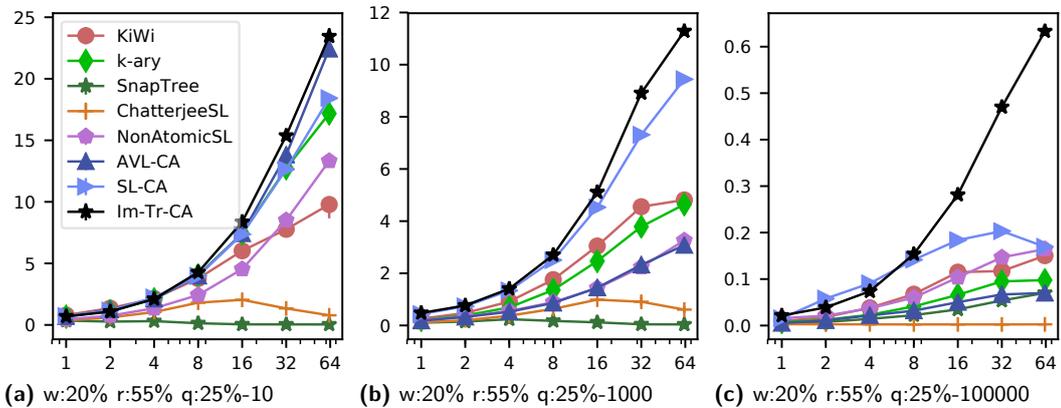

  \centering
  \newcommand{\graphheight}{.22\textheight}
  \subfloat[\textsf{w:20\% r:55\% q:25\%-10}]{
    \includegraphics[height=\graphheight]{{{benchmark_graphs/no+3+10+3+10+ALL+1000000+500000+0.55+0.25+10+0.0+0}}} 
    \label{fig:randr10}
  }
  \subfloat[\textsf{w:20\% r:55\% q:25\%-1000}]{
    \includegraphics[height=\graphheight]{{{benchmark_graphs/no+3+10+3+10+ALL+1000000+500000+0.55+0.25+1000+0.0+0}}}
    \label{fig:randr1000}
  }
  \subfloat[\textsf{w:20\% r:55\% q:25\%-100000}]{
    \includegraphics[height=\graphheight]{{{benchmark_graphs/no+3+10+3+10+ALL+1000000+500000+0.55+0.25+100000+0.0+0}}}
    \label{fig:randr100000}
  }
  \caption{Throughput (operations/$\mu$s) on the y-axis and thread count on the x-axis.}
  \label{fig:randrangeq}
\end{figure} 
\subparagraph{The Random Operations Benchmark} 
This benchmark measures throughput of a mix of operations performed by $N$ threads.
In all captions, benchmark scenarios are described by strings of the form \mbox{\textsf{w:$A$\% r:$B$\% q:$C$\%-$R$}}, meaning that the benchmark performs $(A/2)$\% insert, $(A/2)$\% remove, $B$\% lookup operations and $C$\% range queries of maximum range size $R$.
The range sizes are randomly set to values between 1 and $R$.

\figurename~\ref{fig:randrangeq} shows the results from three scenarios with increasing range sizes.
In the scenario with small range queries of maximum size 10 (\figurename~\ref{fig:randr10}), the best performing data structures (the CA trees and \kary{}) are almost indistinguishable.
\emph{ChatterjeeSL} and \emph{KiWi} that have a global scalability bottleneck, as explained in the previous section, both scale worse in this scenario.
\emph{Im-Tr-CA} scales well in this scenario due to its ability to adapt the sizes of its immutable parts but does not get much benefit from its quick traversal of shared mutable data as conflicts between threads are rare for all data structures in this scenario.

Conflicts are still relatively rare in the scenario with range queries of maximum size 1000 (\figurename~\ref{fig:randr1000}).
The top performing data structures in this scenario (\emph{Im-Tr-CA}, \emph{SL-CA}, \emph{KiWi} and \kary{}) are those with cache locality friendly nodes that store several items in arrays.
However, \emph{Im-Tr-CA} and \emph{SL-CA}, that adapt their synchronization granularity to the scenario at hand, outperform \emph{KiWi} and \kary{} with a wide margin in this scenario.

The scenario that has large range queries of maximum size 100000 (\figurename~\ref{fig:randr100000}) shows the distinguishing feature of \emph{Im-Tr-CA} that outperforms all the other data structures with a large margin.
Conflicts between range queries and update operations are very likely with these large range queries but the conflicts are significantly less costly in \emph{Im-Tr-CA} due to its short critical sections, as is explained in Section~\ref{immopt}.
\subparagraph{Separate Threads for Range Queries and Updates}
Most of the time is spent doing range queries when large range queries are used in the random operations benchmark.
Thus, another benchmark is needed to measure the data structures' ability to handle large range queries concurrently with frequent update operations.
To this end, we use a similar benchmark to the one developed by the \emph{KiWi} authors.
This benchmark is motivated by large scale applications that require quick updates of a data set while other threads do large linearizable range queries concurrently (for analytics)~\cite{BasinKiWi}.
In this benchmark, half the threads do update operations (insert and remove with equal probability) while the other half do range queries with a range of fixed size.
The throughput for updates is presented separately from the range query throughput so that one can study the performance of these operations separately.
Note that in the graphs that show the range query throughput, the number of operations per $\mu$s multiplied by the range query size is shown on the y-axis to make the graphs more readable.

\begin{figure}[t]
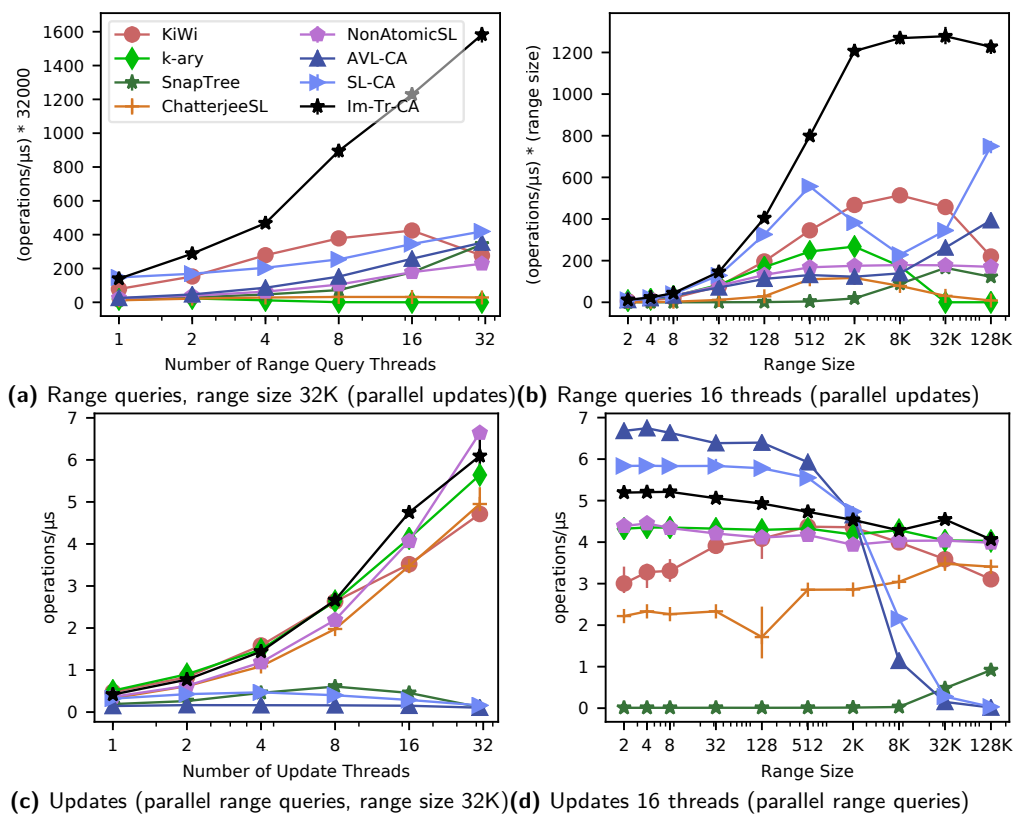
 
  \centering
  \newcommand{\graphheight}{.22\textheight}
  \subfloat[\textsf{Range queries, range size 32K (parallel updates)}]{ 
    \includegraphics[height=\graphheight]{{{benchmark_graphs/no+3+10+3+10+ALL+1000000+500000+32000_range}}} 
    \label{fig:threadsseprange}
  }
  \subfloat[\textsf{Range queries 16 threads (parallel updates)}]{
    \includegraphics[height=\graphheight]{{{benchmark_graphs/no+threads32_1000000_range}}} 
    \label{fig:rangeseprange}
  } \\[-1em]  
  \subfloat[\textsf{Updates (parallel range queries, range size 32K)}]{
    \includegraphics[height=\graphheight]{{{benchmark_graphs/no+3+10+3+10+ALL+1000000+500000+32000_put_marg}}} 
    \label{fig:threadssepput}
  }   
  \subfloat[\textsf{Updates 16 threads (parallel range queries)}]{
    \includegraphics[height=\graphheight]{{{benchmark_graphs/no+threads32_1000000_put_marg}}} 
    \label{fig:rangesepput}
  }
  \caption{Results for benchmark with separate threads doing updates and range queries.}
  \label{fig:sepupdaterangeqtheads}
\end{figure}
\figurename~\ref{fig:threadsseprange} and \figurename~\ref{fig:threadssepput} show the results of this benchmark with a range query size of 32K and with varying thread counts.
In \figurename~\ref{fig:rangeseprange} and \figurename~\ref{fig:rangesepput}, the thread count is fixed to 32 (16 updaters and 16 threads doing range queries) and the x-axis shows varying range query sizes.
First of all, \emph{Im-Tr-CA} with its short range query critical section is overall the fastest data structure in the scenarios.
\emph{KiWi} is the second most performant data structure in the scenarios with range query sizes larger than 2000.
The bumpy performance of \emph{CA-SL} in \figurename~\ref{fig:rangeseprange} can be explained by the fact that the range queries acquire the base node locks in read-mode which enables concurrent range queries to bypass waiting update operations and take over the lock.
\emph{CA-SL} thus provides good throughput for range queries with a range size of 512 because conflicts are rare and with a range size of 128K as ``conflicts'' with other range queries that have already acquired the relevant base nodes are common. %
\emph{ChatterjeeSL}'s and \emph{KiWi}'s update operations need to read the RangeCollector list (\emph{ChatterjeeSL}) or the global version number (\emph{KiWi}) which are updated by range queries.
The more frequent updates of these global objects with smaller range queries can explain the slight partial upward trend that exists for \emph{ChatterjeeSL} and \emph{KiWi} in Figure~\ref{fig:rangesepput}.
\kary{}'s range queries are starved by update operations in the scenarios with large range queries.
The \emph{SnapTree}'s operations are slow in most scenarios due to its coarse-grained approach for doing range queries, but the \emph{SnapTree}'s performance is better in scenarios with larger and less frequent range queries.

\tablename~\ref{tab:nodestats} shows the average number of base nodes and the number of base nodes traversed per range query in \emph{Im-Tr-CA} after the benchmark runs of the scenarios displayed in \figurename~\ref{fig:rangeseprange} and \figurename~\ref{fig:rangesepput}.
It is evident from the table data that \emph{Im-Tr-CA}'s range queries spend a very short time traversing shared mutable data even for large range queries (e.g. approximately the time it takes to traverse 13 base nodes in the case with range queries of size 32K).
After the base nodes have been traversed, the collected immutable data can be traversed without any need to care about other threads and without disturbing other threads.

 \begin{table}[h]
  \caption{Statistics for \emph{Im-Tr-CA} in the scenarios displayed in \figurename~\ref{fig:rangeseprange} and \figurename~\ref{fig:rangesepput}.\label{tab:nodestats}}
      \begin{tabular}{c|cccccccccc}
      Range Size  & 2 & 4 & 8 & 32 & 128 & 512 & 2K & 8K & 32K & 128K \\\hline
      \# base nodes & 2.5K & 2.1K & 1.7K & 1.0K & 590 & 390 & 310 & 310 & 390 & 430 \\ 
      $\frac{\text{\# traversed base nodes}}{\text{\# range queries}}$ & 1.0 & 1.0 & 1.0 & 1.0 & 1.1 & 1.2 & 1.6 & 3.5 & 13 & 56 \\
    \end{tabular}
\end{table} 
 \section{Conclusion\label{sec:conc}}
 A new CA tree variant that makes use of an immutable data structure has been presented.
 The advantage of the new CA tree variant over the CA tree variants that use mutable data structures as the sequential data structure component is that the new variant drastically reduces the time period in which conflicts between large range queries and other operations can happen.
 Compared to all other data structures with linearizable range query support, the CA trees have the advantage that they dynamically adapt the synchronization granularity to fit the workload at hand.
 The experimental comparison shows that the presented implementation's quick traversal of shared mutable data and cache friendly design  makes the implementation outperform the best of the other data structures with a wide margin in scenarios with large range queries.
 Furthermore, the new CA tree variant also performs better or close to the best of the other data structures in scenarios with small range queries due to its ability to dynamically change its synchronization granularity.  
 As future work, we plan to design and evaluate a lock-free CA tree variant.
 A lock-free CA tree variant could potentially give even better performance as it could avoid priority inversions and other lock related problems.

\subparagraph*{Acknowledgments}
Vincent Gramoli gave me the idea of looking into immutable data structures in combination with the CA tree.
Amelie Lind, Martin Viklund, Stephan Brandauer, Andreas Löscher, Elias Castegren and Konstantinos Sagonas helped me improve the language.
This work was supported by the Linnaeus centre of excellence UPMARC (Uppsala Programming for Multicore
Architectures Research Center).

\appendix
\section{Source Code}
The source code for the CA tree implementations and the benchmarks can be found online (\url{https://www.it.uu.se/research/group/languages/software/im_tr_ca}).

\section{Results with Other Set Sizes}
Results corresponding to the results in figures \ref{fig:randrangeq} and \ref{fig:sepupdaterangeqtheads} but with smaller set sizes (the key range size $S=10^5$) can be found in figures \ref{fig:randrangeq100000} and \ref{fig:sepupdaterangeqtheads100000}.
The corresponding results for larger set sizes ($S=10^7$) can be found in figures \ref{fig:randrangeq10000000} and \ref{fig:sepupdaterangeqtheads10000000}.
The statistics corresponding to the statistics in \tablename~\ref{tab:nodestats} but with the smaller and larger set sizes can be found in tables~\ref{tab:nodestats100000} and \ref{tab:nodestats10000000}.

In the cases with the smallest set size ($S=10^5$), the ranges of size 32K and 128K span 32\% and 100\% of the set represented by the data structures; see \figurename~\ref{fig:sepupdaterangeqtheads100000}.
\emph{Im-Tr-CA}'s range queries lock out update operations from the portion of the set that is covered by the range query.
Even though this only happens for a short period of time as Table~\ref{tab:nodestats100000} shows, it still has a negative effect on update operations as \figurename~\ref{fig:rangesepput100000} shows.
This is compensated by \emph{Im-Tr-CA}'s excellent performance for range queries in these scenarios as \figurename~\ref{fig:rangeseprange100000} shows.

In the cases with the largest set size ($S=10^7$), the ranges span a smaller part of the sets represented by the data structures which explains why many of the other data structures are closer to \emph{Im-Tr-CA} in these scenarios; see figures~\ref{fig:randrangeq10000000} and \ref{fig:sepupdaterangeqtheads10000000}.

\bibliography{biblio}

\begin{figure}[b]
  \centering
  \newcommand{\graphheight}{.22\textheight}
  \subfloat[\textsf{w:20\% r:55\% q:25\%-10}]{ 
    \includegraphics[width=0.31\textwidth]{{{benchmark_graphs/no+3+10+3+10+ALL+100000+50000+0.55+0.25+10+0.0+0}}} 
    \label{fig:randr10:100000}
  }
  \subfloat[\textsf{w:20\% r:55\% q:25\%-1000}]{
    \includegraphics[width=0.325\textwidth]{{{benchmark_graphs/no+3+10+3+10+ALL+100000+50000+0.55+0.25+1000+0.0+0}}}
    \label{fig:randr1000:100000}
  }
  \subfloat[\textsf{w:20\% r:55\% q:25\%-100000}]{
    \includegraphics[width=0.315\textwidth]{{{benchmark_graphs/no+3+10+3+10+ALL+100000+50000+0.55+0.25+100000+0.0+0}}}
    \label{fig:randr100000:100000}
  }
  \caption{Results with key range of size $10^5$ which corresponds to a set size of approximately $5\times 10^4$. Throughput (operations/$\mu$s) on the y-axis and thread count on the x-axis.}
  \label{fig:randrangeq100000}
\end{figure} 
\begin{figure}[t] 
  \centering
  \newcommand{\graphheight}{.22\textheight}
  \subfloat[\textsf{Range queries, range size 32K (parallel updates)}]{ 
    \includegraphics[height=\graphheight]{{{benchmark_graphs/no+3+10+3+10+ALL+100000+50000+32000_range}}} 
    \label{fig:threadsseprange100000}
  }
  \subfloat[\textsf{Range queries 16 threads (parallel updates)}]{
    \includegraphics[height=\graphheight]{{{benchmark_graphs/no+threads32_100000_range}}} 
    \label{fig:rangeseprange100000}
  } \\[-1em]  
  \subfloat[\textsf{Updates (parallel range queries, range size 32K)}]{
    \includegraphics[height=\graphheight]{{{benchmark_graphs/no+3+10+3+10+ALL+100000+50000+32000_put_marg}}} 
    \label{fig:threadssepput100000}
  }      
  \subfloat[\textsf{Updates 16 threads (parallel range queries)}]{
    \includegraphics[height=\graphheight]{{{benchmark_graphs/no+threads32_100000_put_marg}}} 
    \label{fig:rangesepput100000}
  }
  \caption{Results for benchmark with separate threads doing updates and range queries with key range of size $10^5$ which corresponds to a set size of approximately $5\times 10^4$.}
  \label{fig:sepupdaterangeqtheads100000}
\end{figure} 
 \begin{table}[h]\footnotesize  
  \caption{Statistics for \emph{Im-Tr-CA} in the scenarios displayed in \figurename~\ref{fig:rangeseprange100000} and \figurename~\ref{fig:rangesepput100000}.\label{tab:nodestats100000}}
      \begin{tabular}{c|cccccccccc} %
      Range Size  & 2 & 4 & 8 & 32 & 128 & 512 & 2K & 8K & 32K & 128K \\\hline
      \# base nodes & 890 & 720 & 570 & 330 & 190 & 120 & 97 & 130 & 90 & 96 \\ 
      $\frac{\text{\# traversed base nodes}}{\text{\# range queries}}$ & 1.0 & 1.0 & 1.1 & 1.1 & 1.2 & 1.6 & 3.0 & 11 & 27 & 98 \\
    \end{tabular}
 \end{table}

\begin{figure}[b]
  \centering
  \newcommand{\graphheight}{.22\textheight}
  \subfloat[\textsf{w:20\% r:55\% q:25\%-10}]{
    \includegraphics[height=\graphheight]{{{benchmark_graphs/no+3+10+3+10+ALL+10000000+5000000+0.55+0.25+10+0.0+0}}} 
    \label{fig:randr10:10000000}
  }
  \subfloat[\textsf{w:20\% r:55\% q:25\%-1000}]{
    \includegraphics[height=\graphheight]{{{benchmark_graphs/no+3+10+3+10+ALL+10000000+5000000+0.55+0.25+1000+0.0+0}}}
    \label{fig:randr1000:10000000}
  }
  \subfloat[\textsf{w:20\% r:55\% q:25\%-100000}]{
    \includegraphics[height=\graphheight]{{{benchmark_graphs/no+3+10+3+10+ALL+10000000+5000000+0.55+0.25+100000+0.0+0}}}
    \label{fig:randr100000:10000000}
  }
  \caption{Results with key range of size $10^7$ which corresponds to a set size of approximately $5 \times 10^6$. Throughput (operations/$\mu$s) on the y-axis and thread count on the x-axis.}
  \label{fig:randrangeq10000000}
\end{figure} 
\begin{figure}[t] 
  \centering
  \newcommand{\graphheight}{.22\textheight}
  \subfloat[\textsf{Range queries, range size 32K (parallel updates)}]{ 
    \includegraphics[height=\graphheight]{{{benchmark_graphs/no+3+10+3+10+ALL+10000000+5000000+32000_range}}} 
    \label{fig:threadsseprange10000000}
  }
  \subfloat[\textsf{Range queries 16 threads (parallel updates)}]{
    \includegraphics[height=\graphheight]{{{benchmark_graphs/no+threads32_10000000_range}}} 
    \label{fig:rangeseprange10000000}
  } \\[-1em]  
  \subfloat[\textsf{Updates (parallel range queries, range size 32K)}]{
    \includegraphics[height=\graphheight]{{{benchmark_graphs/no+3+10+3+10+ALL+10000000+5000000+32000_put_marg}}} 
    \label{fig:threadssepput10000000}
  }      
  \subfloat[\textsf{Updates 16 threads (parallel range queries)}]{
    \includegraphics[height=\graphheight]{{{benchmark_graphs/no+threads32_10000000_put_marg}}} 
    \label{fig:rangesepput10000000}
  }
  \caption{Results for benchmark with separate threads doing updates and range queries with key range of size $10^7$ which corresponds to a set size of approximately $5 \times 10^6$.}
  \label{fig:sepupdaterangeqtheads10000000}
\end{figure}
 \begin{table}[h]\footnotesize  
  \caption{Statistics for \emph{Im-Tr-CA} in the scenarios displayed in \figurename~\ref{fig:rangeseprange10000000} and \figurename~\ref{fig:rangesepput10000000}.\label{tab:nodestats10000000}}
      \begin{tabular}{c|cccccccccc} %
      Range Size  & 2 & 4 & 8 & 32 & 128 & 512 & 2K & 8K & 32K & 128K \\\hline
      \# base nodes & 6.0K & 5.3K & 4.5K & 2.9K & 1.8K & 1.2K & 900 & 860 & 1.0K & 1.1K \\ 
      $\frac{\text{\# traversed base nodes}}{\text{\# range queries}}$ & 1.0 & 1.0 & 1.0 & 1.0 & 1.0 & 1.1 & 1.2 & 1.7 & 4.2 & 15 \\ %
    \end{tabular}
 \end{table}

\end{document}